\documentclass[a4paper,11pt]{article}
\pdfoutput=1 
\usepackage{jinstpub} 
\usepackage{lineno}
\usepackage{natbib}
\usepackage[normalem]{ulem}
\usepackage{multicol}
 \usepackage{multirow}
 \def\Plus{\texttt{+}}
\def\Minus{\texttt{-}}
\usepackage{adjustbox}
\title{\boldmath J-PET detection modules based on plastic scintillators for performing studies with positron and positronium beams}
\author[1,2,3,*]{S. Sharma,\note{Corresponding author.}}
\author[1,2,3]{J. Baran,}
\author[4,5]{R.S. Brusa,}
\author[5]{R. Caravita,}
\author[1,2,3]{N. Chug,}
\author[1,2,3]{A. Coussat,}
\author[6]{C. Curceanu,}
\author[1,2,3]{E. Czerwiński,}
\author[1,2,3]{M. Dadgar,}
\author[1,2,3]{K. Dulski,}
\author[1,2,3]{K. Eliyan,}
\author[1,2,3]{A. Gajos,}
\author[7]{B.C. Hiesmayr,}
\author[1,2,3]{K. Kacprzak,}
\author[1,2,3]{\L. Kap{\l}on,}
\author[8]{K. Klimaszewski,}
\author[9]{P. Konieczka,}
\author[1,2]{G. Korcyl,}
\author[1]{T. Kozik,}
\author[9]{W. Krzemień,}
\author[1,2,3]{D. Kumar,}
\author[4,5]{S. Mariazzi,}
\author[1,2,3]{S. Niedźwiecki,}
\author[4,5]{L. Panasa,}
\author[1,2,3]{S. Parzych,}
\author[4,5]{L. Povolo,}
\author[1,2,3]{E. Perez del Rio,}
\author[9]{L. Raczyński,}
\author[1,2,3]{Shivani,}
\author[9]{R.Y. Shopa,}
\author[1,2,3]{M. Skurzok,}
\author[1,2,3]{E.Ł. Stępień,}
\author[1,2,3]{F. Tayefi,}
\author[1,2,3]{K. Tayefi,}
\author[9]{W. Wiślicki,}
\author[1,2,3]{P. Moskal,}
\affiliation[1]{Faculty of Physics, Astronomy and Applied Computer Science, Jagiellonian University, Krakow, Poland}
\affiliation[2]{Total-Body Jagiellonian-PET Laboratory, Jagiellonian University, Kraków, Poland}
\affiliation[3]{Center for Theranostics, Jagiellonian University, Cracow, Poland}
\affiliation[4]{Department of Physics, University of Trento, via Sommarive 14, 38123 Povo,Trento, Italy}
\affiliation[5]{TIFPA/INFN, via Sommarive 14, 38123 Povo,Trento, Trento, Italy}
\affiliation[6]{INFN, Laboratori Nazionali di Frascati, Frascati, Italy}
\affiliation[7]{Faculty of Physics, University of Vienna, Vienna, Austria}
\affiliation[8]{Department of Complex Systems, National Centre for Nuclear Research, Otwock-Świerk, Poland}
\affiliation[9]{High Energy Physics Division, National Centre for Nuclear Research, Otwock-Świerk, Poland}

\emailAdd{sushil.sharma@uj.edu.pl}
\abstract{The J-PET detector, which consists of inexpensive plastic scintillators, has demonstrated its potential in the study of fundamental physics. In recent years, a prototype with 192 plastic scintillators arranged in 3 layers has been optimized for the study of positronium decays. This allows performing precision tests of discrete symmetries (C, P, T) in the decays of positronium atoms. Moreover, thanks to the possibility of measuring the polarization direction of the photon based on Compton scattering, the predicted entanglement between the linear polarization of annihilation photons in positronium decays can also be studied. Recently, a new J-PET prototype was commissioned, based on a modular design of detection units. Each module consists of 13 plastic scintillators and can be used as a stand-alone, compact and portable detection unit. In this paper, the main features of the J-PET detector, the modular prototype and their applications for possible studies with positron and positronium beams are discussed. Preliminary results of the first test experiment performed on two detection units in the continuous positron beam recently developed at the Antimatter Laboratory (AML) of Trento are also reported.}
\proceeding{Proc. of the 23rd International Workshop on Radiation Imaging Detectors\\}
\begin{document}
\maketitle
\flushbottom
\section{Introduction}
\label{sec:intro}
\textbf{J}agiellonian \textbf{P}ositron \textbf{E}mission \textbf{T}omograph (J-PET)  is the first tomograph based on the idea of using plastic scintillators instead of crystals as currently used in commercial tomographs~\cite{MOS11,MOS12}. Plastic scintillators have excellent time resolution and are therefore good candidates for building TOF-PET tomographs~\cite{MOS12B}. The novelty that distinguishes J-PET from other tomographs is its potential to conduct studies of fundamental physics problems~\cite{MOS16} and positronium imaging~\cite{MOS19,MOS21A,MOSPOS}. Therefore, it can be used in a dual role, both as a PET scanner and as a multimodule detector. The J-PET detector is optimised for studying the decay of positronium atoms (Ps, the bound state of electron ($e^-$) and positron ($e^+$))~\cite{MOS20,DUL21}. In recent years, data have been collected with the 3-layer prototype of J-PET~\cite{NIE17}, which consists of 192 detection modules, demonstrating its applicability not only in medical physics~\cite{MOS21A,RAC20,SHO21} but also in the study of fundamental physics~\cite{MOS21B}. Following the J-PET technology, a new prototype was recently built based on a modular design consisting of 24 individual detection units~\cite{MOS21C}. Thanks to the modular design, the detection units can be conveniently transported to other research facilities to perform experiments. At the University of Trento, a continuous positron beam has been put into operation in the Antimatter Laboratory (AML). With the know-how to fabricate efficient positron/positronium converters~\cite{MAR10,MAR21} and to manipulate positronium atoms into a metastable state with increased lifetime~\cite{AMS19}, the generation of Ps beams is envisaged~\cite{MAR20A}. Two detection modules with a supported data acquisition chain have recently been moved to the AML to perform studies with positrons and positronium beams to investigate fundamental physics not yet explored. The outline of the draft is divided as follows. In the next section, an introduction to the 3-layer protoype of J-PET is given, followed by the studies that are currently being performed. Then, the modular detection units are briefly described and their possible applications with $e^+$ and Ps beams are discussed. Finally, the preliminary results of the first experiment using a continuous positron beam to reconstruct the beam spot with two detection modules are reported.    
\section{J-PET tomograph as multi-photon detector}
\label{sec:jpet}
The 3-layer prototype is developed to test the concept of constructing a cost-effective total-body PET from plastic scintillators~\cite{NIE17,MOS16B,MOS21C}. To obtain a longer axial field of view (AFOV), strips of 500$\times$19$\times$7~mm$^3$ plastic scintillators are used. In the design of J-PET, 192 plastic scintillators (EJ-230) are arranged in 3 concentric cylinders with radii 42.5~cm, 46.75~cm, and 57.5~cm, respectively. Signals from each scintillator are read out with an R9800 Hamamatsu photomultiplier on each side. Data are measured and stored in triggerless mode, which can handle data streams of up to 8 Gbps~\cite{PAL17,KOR18}. To take advantage of the excellent time resolution, Time Over Threshold (TOT) is measured instead of the charge collection. The energy deposition in a given interaction of photons within the scintillator is estimated based on the established relationship between TOT and the energy deposition~\cite{SHA20B}. A special framework based on advanced C++ routines and ROOT (Data Analysis Framework from CERN) was also developed to analyse the measured data~\cite{KRZ20}. Hit position and hit time of a photon interaction (hit) within the scintillator are calculated based on the measured light signals read from both ends of a scintillator~\cite{MOS14}. The hit time is calculated as the average of the times of the light signals arriving at both ends, while the hit position is calculated as the product of the difference in the arrival times of the light signal at both ends multiplied by half of their effective speed~\cite{MOS15,RAC17}.   
\begin{figure}[htbp]
\centering 
\includegraphics*[width=.8\textwidth,origin=c,angle=0]{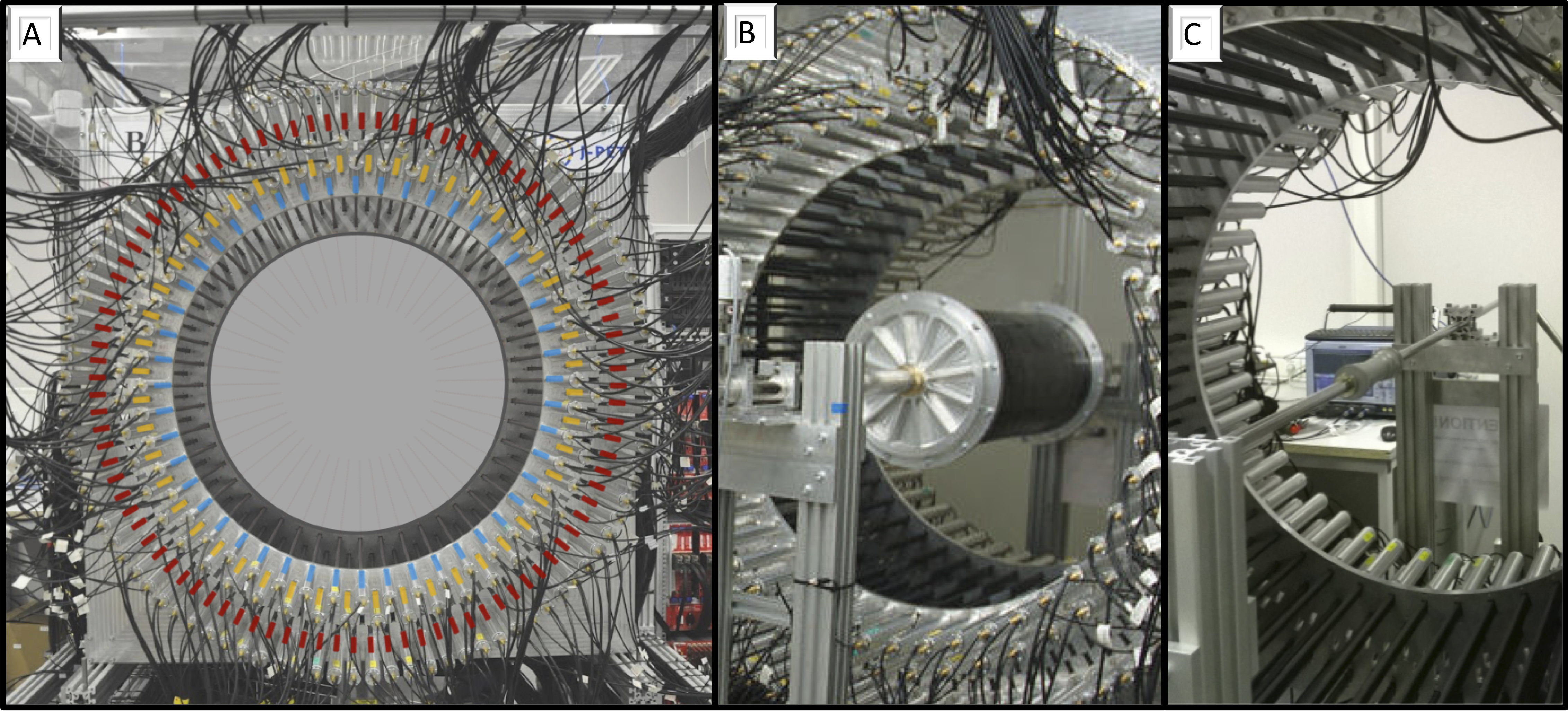}
\caption{\label{fig:jpet}(A) Shows an image of J-PET in the laboratory. 3-Layers (blue, yellow, red) represent the angular arrangement of the scintillator strips and their cross-section in the plane. (B) shows the installation of hollow cylindrical chamber in the centre of J-PET, while (C) represents the small aluminium chamber.}
\end{figure}
Fig.~\ref{fig:jpet} shows the pictures of the J-PET detector (left), the installation of the hollow cylindrical chamber (centre), and the small aluminium chamber (right). Before turning to the modular prototype of J-PET, the next section briefly discusses the physics aspects of the J-PET detector in studying the decays of positronium atoms (Ps)~\cite{MOS16}.
\subsection{Fundamental studies in decays of Ps using J-PET}
Ps, being a pure leptonic system of electron and positron, is an excellent probe to test the bound state of quantum electrodynamics~\cite{BAS19}. Ps can be formed in one of two ground states: Singlet state ($^1$S$_0$: para-positronium (p-Ps)) with lifetime 125~ps or in the triplet state ($^3$S$_1$: ortho-positronium(o-Ps)) with lifetime 142~ns. Under the charge conjugation condition, p-Ps and o-Ps decay into an even number (2n$\gamma$; n=1,2,.), while o-Ps decays into an odd number ((2n+1)$\gamma$; n=1,2,.) of photons, predominantly 2$\gamma$ and 3$\gamma$, respectively. In studying the decays of Ps atoms, several fundamental problems can be investigated, e.g., quantum entanglement, violation of discrete symmetries, etc. We briefly discuss here the studies that are currently being carried out with the J-PET detector.\\
\textbf{Photon polarization and quantum entanglement:}
\label{sec:ent}
The measurement of entanglement in the annihilation photons emitted in Ps is one of the important aspects that can be studied with the J-PET detector. It is predicted that the linear polarizations of back-to-back photons emitted in the singlet state (p-Ps) of Ps are orthogonally correlated~\cite{WHE46,WARD47}. Measurement of the correlation between the polarizations of annihilation photons can be used to observe the entangled state~\cite{BOH57,CAR19,HIE17,BEA19}, a subject of fundamental importance that has direct application in medical imaging~\cite{TOG16,MOS22}. There are no mechanical polarizers to measure the polarization of 511 keV photons. However, Compton scattering can be used as a polarizer for such measurements~\cite{NIS29}. J-PET is capable of registering both annihilation photons before and after their scattering with angular resolution$\approx$1$^o$~\cite{KAM16}. Based on the fact that in Compton scattering, a photon is scattered most likely at right angles to the direction of linear polarization of the incident photon, the polarization of the photon can be defined as \textbf{$\vec{\epsilon}=\vec{k}\times\vec{k^{'}}$}~\cite{MOS16,MOS18}. By measuring the polarization of each photon, one can measure the azimuthal correlation between their polarizations and test the theoretically predicted claims for entanglement~\cite{BOH57,CAR19,HIE17,BEA19}. Entanglement studies can be extended to the o-Ps $\rightarrow$3$\gamma$ case~\cite{BEA19}.\\
\textbf{Symmetry violation - test of discrete symmetries in decays of Ps atoms:} Ps exhibits interesting properties that make it an exotic atom for performing the tests on discrete symmetries. For example, it is an eigenstate of the parity operator (P) since it is bound by a central potential. Moreover, Ps is a system of a particle and its antiparticle that remains symmetric in their exchange and thus it is also an eigenstate of the charge conjugation operator C. Therefore, positronium is also an eigenstate of the CP operator. In conjunction with the CPT theorem, one can test the T violation effects separately or in combination for CPT test. Therefore, Ps can serve as an excellent laboratory to perform tests for C, P, CP and CPT violations. In 1988, studying the o-Ps$\rightarrow$3$\gamma$ decays, Bernreuther and coworkers suggested that to test the discrete symmetries, odd-symmetry operators can be constructed using the spin of the o-Ps atom and momenta of annihilation photons~\cite{BER88}. Non vanishing expectation value of these operators will be the confirmation of the symmetry violation. Table~\ref{tab:i} represents the list of operators for the the discrete symmetries test~\cite{MOS16}. Minus sign represents the odd-symmetric operators which are sensitive to observe the violation effects for discrete symmetries. In addition, with the ability of J-PET detector to measure the polarization direction of photons~\cite{MOS18}, additional operators are also constructed utilizing the photon's polarization direction which are unique and currently possible only with the J-PET~\cite{MOS16}.
\begin{table}[htbp]
\centering
\caption{\label{tab:i} Odd-symmetry operators constructed of the photons momenta ($\vec{k}_i$) and polarization ($\vec{\epsilon}_i$), and spin of o-Ps ($\vec{S}$)~\cite{MOS16}}
\begin{tabular}{|c|c c c c c||c|c c c c c|}
\hline Odd symmetric & \textbf{C} & \textbf{P} & \textbf{T} & \textbf{CP} & \textbf{CPT} & Odd symmetric  & \textbf{C} & \textbf{P} & \textbf{T} & \textbf{CP} & \textbf{CPT} \\
\hline
$\vec{S}$ . $\vec{k}_{\,1}$ & \Plus & \textbf{\Minus} & \Plus & \textbf{\Minus} & \textbf{\Minus} & $\vec{k}_{\,1}$ . $\vec{\epsilon}_{\,2}$ & \Plus & \textbf{\Minus} & \textbf{\Minus} & \textbf{\Minus} & \Plus \\
\hline
$\vec{S}$ . ($\vec{k}_{\,1} \times \vec{k}_{\,2}$) & \Plus & \Plus & \textbf{\Minus} & \Plus &  \textbf{\Minus} & $\vec{S}$ . $\vec{\epsilon}_{\,1}$ & \Plus & \Plus & \textbf{\Minus} & \Plus & \textbf{\Minus}\\
\hline
$(\vec{S}$ . $\vec{k}_{\,1})$$(\vec{S}$ . ($\vec{k}_{\,1} \times \vec{k}_{\,2})$ & \Plus & \textbf{\Minus} & \textbf{\Minus} &\textbf{\Minus} & \Plus & $\vec{S}$ . ($\vec{k}_{\,2} \times \vec{\epsilon}_{\,1})$ & \Plus & \textbf{\Minus} & \Plus & \textbf{\Minus} & \textbf{\Minus} \\
\hline
\end{tabular}
\end{table}

In table~\ref{tab:i}, one can see that none of the operators is sensitive to the C symmetry test. However, tests for violation of C symmetry can be performed by examining C-prohibited Ps-decays (e.g. p-Ps$\rightarrow$3$\gamma$, o-Ps$\rightarrow$2$\gamma$, 4$\gamma$,.). In symmetry studies with J-PET, the estimate of the o-Ps spin is based on the intrinsic polarization of positrons emitted in $\beta^+$ decays~\cite{MOS21B}. These are longitudinally polarized due to parity violation, which is proportional to the velocity of the positrons. Different types of chambers can be used depending on the specifics of the operators. For the operators involving the spin of o-Ps, large hollow chambers (see Fig.~\ref{fig:jpet} (B)) are used, the inner wall of which are coated with porous materials to increase the probability of Ps formation. The annihilation points of the o-Ps in the chamber wall can be reconstructed using the trilateration method~\cite{GAJ16}, which gives the direction of the positrons with respect to the known position of the source and thus allows the o-Ps spin to be estimated~\cite{MOS16,MOS21B}. Small chambers (Fig.~\ref{fig:jpet} (C)) are used for the other operators, in particular with photon polarization.
\section{Modular J-PET and its possible applications with positron and positronium beams}
The modular J-PET is a new prototype, which consists of 24 independent detection modules. Each module consists of 13 plastic scintillators of size 500$\times$24$\times$6 mm$^3$, read out at each end by a SiPM matrix, together with their front-end electronics housed in a single module. Thanks to their modular design and FPGA-based compact data acquisition, they can be easily transported to be used as potential detectors in different laboratories. In this context, we have explored the possibility of using the modular detection units for studies with positron and positronium beams at the AML in Trento~\cite{POV}.
A continuous positron beam was recently commissioned at the AML. In the future, the continuous beam will be injected into a Surko trap~\cite{DAN15} where positrons will be trapped, stored for fraction of seconds and then bunched to form pulses containing up to 10$^4$ positrons. Implantation of positron pulses in efficient positron/positronium converters~\cite{MAR10,LIS12,MAR21D} allows producing dense Ps clouds~\cite{CAS06,AGH15}. In particular, in recent years the possibility to populate the long-lived 2$^3$S state of positronium via spontaneous~\cite{AMS19} and stimulated~\cite{ANT19} decay from the 3$^3$P level (previously reached via 1$^3$S$\rightarrow$3$^3$P laser excitation) has been demonstrated~\cite{AGH16}. A monochromatic pulsed 2$^3$S positronium beam with low angular divergence can then be produced by placing an iris diaphragm in front of the target~\cite{MAR20A}. 
By employing properly polarized laser pulses, the production of Ps in 2$^3$S with fully controlled quantum numbers looks feasible. In studying the annihilation of Ps in 3-photons, an interesting fundamental problem, the experimental measurement of the quantum entanglement of the polarization of the annihilation photons could be addressed for the first time. Theoretical studies predict that the entanglement type of the 3-photons depends on the quantum numbers of the annihilating positronium~\cite{BEA19}. 
Ps in the 2$^3$S state is also of interest for direct measurements of gravitational interaction on antimatter~\cite{OBE02,MAR20}. Indeed, Ps excited in a long-lived state~\cite{MIL02} together with antihydrogen~\cite{AMS21,ALP,GBAR} and muonium~\cite{ANT18}, have been proposed as a probe for test of weak equivalence principle on antimatter. A possible experimental scheme consists in the employment of a  Ps
beam in the metastable 2$^3$S state crossing a deflectometer or an interferometer to form a fringe pattern~\cite{OBE02,MAR20}. In presence of an external force, the fringe pattern shows a displacement that is proportional to the acceleration experienced by the Ps~\cite{MAR20}. In order to detect such a fringe pattern shift, Ps atom distribution on a plane could be scanned by using a slit or a material grating~\cite{MAR20}. Ps annihilating on the obstacles and the ones crossing it can then be counted as a function of the position of the slit/grating and the Ps spatial distribution on the plane can be reconstructed.
A detector able to resolve the annihilation points of Ps along the beam direction (to distinguish the annihilations on the obstacles from the ones occurring forward) is needed.
To verify the applicability of the J-PET detection units for this purpose, two such units with complete readout electronics were transported to AML. A test run was performed to measure the spatial resolution that can be achieved with only 2 detection units with e$^+$ beam. The details and first results of the test can be found in the next section.
\section{Performance study of two J-PET detection units with positron beam}
\begin{figure}[htbp]
\centering  
    \includegraphics*[width=.99\textwidth,origin=c,angle=0]{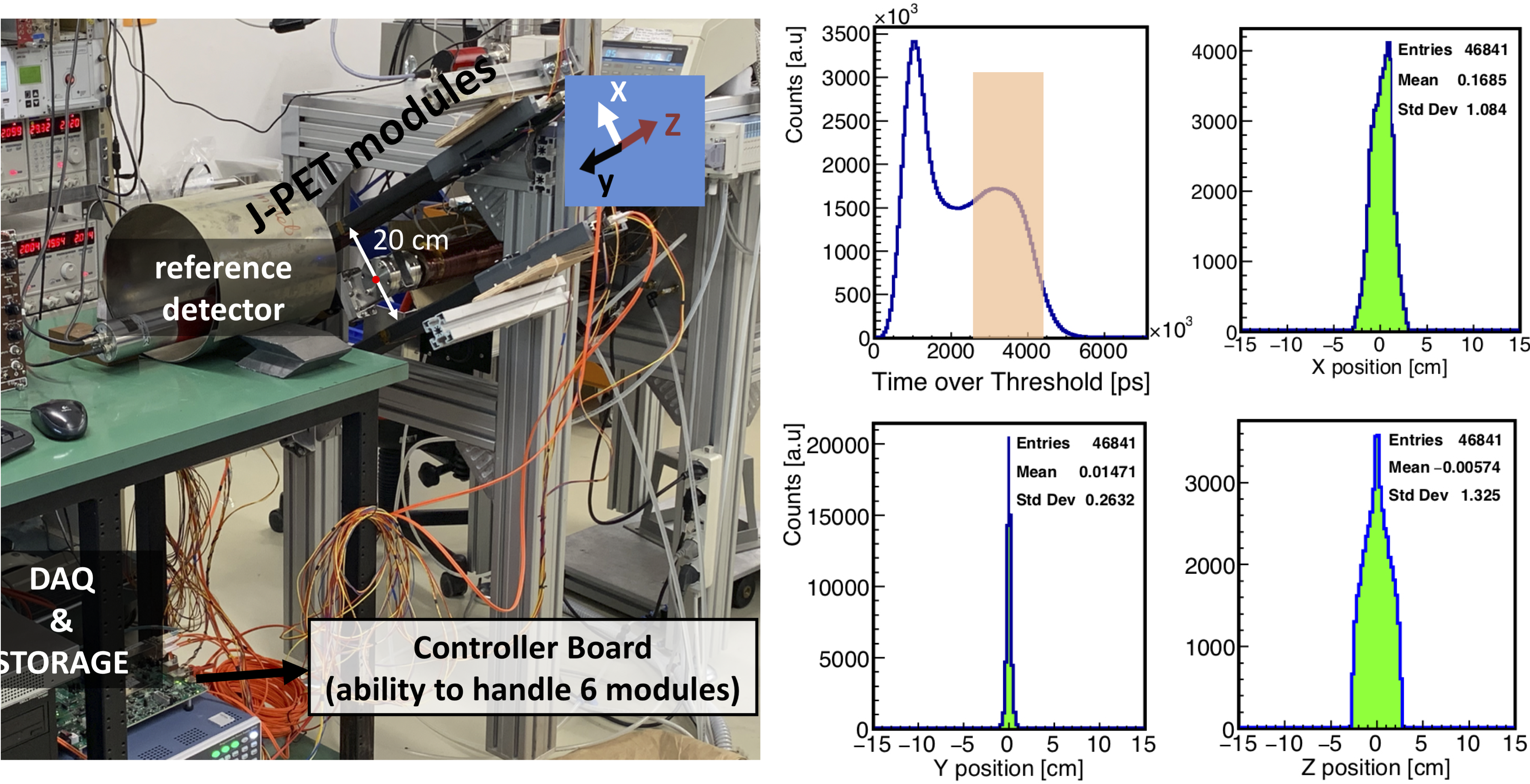}
\caption{\label{fig:setups}The photo of the experimental setup (left). Two modules are 20 cm apart and centered around the e$^+$ annihilation points. The X,Y,Z directional frame ( width(24~mm), thickness(6~mm), length(500~mm) ) of the modules is such that the Y-axis is along the direction of the beam, while the X- and Z-directions are perpendicular and parallel to the plane of the J- PET module, respectively. On the right is the preliminary measured TOT distribution (upper left inset) and the 3D (X,Y,Z) projections of the reconstructed vertices.}
\end{figure}

To investigate the performance of the J-PET modules, 511~keV photons emitted by the annihilations of $e^+$ implanted with the AML beam into a stainless-steel flange have been recorded. Two modules were placed 20~cm apart from the e$^+$ beam spot (red dot) as shown in the left panel of Fig.~\ref{fig:setups}. Binary data registered by the FPGA cards were processed using framework software developed by the J-PET collaboration \cite{KRZ20}. Hit time and hit position are reconstructed as described above. Signals from each SiPM are sampled at two thresholds in the voltage domain (30~mV and 70~mV). TOT as a measure of the energy deposition by photons interacting in a scintillator (hit) is calculated as the sum of the TOTs at both thresholds of the connected SiPMs. In the right panel of Fig.~\ref{fig:setups}, the upper left inset shows the measured TOT spectra. 
Since TOT is the measure of energy deposition, higher TOT values are expected with increasing energy deposition~\cite{SHA20B}. The structure with two peaks corresponds to the energy depositions by 511 keV photons and their scattered photons. The first peak results from the interactions of the scattered photons, while the second enhancement indicated in orange color represents the contribution by the 511~keV photons. In the analysis of events with 2 hits by 511 keV photons, the annihilation points of $e^+$ are reconstructed. For the selection of 511 keV interactions, the first criterion is based on the measured TOT values for both hits. Only those events for which the TOT values are lying in the shaded region (orange) were selected. The second criterion is based on their angular correlation, i.e., the photons that caused 2 hits are considered only if emitted in back-to-back directions. After the selection of the 511~keV photons, the annihilation vertices are reconstructed. The projections of the reconstructed vertices on each axis are shown in the upper right and lower insets of Fig.~\ref{fig:setups}.
Preliminary results of the analysis performed over a set of data measured with a stainless-steel flange are presented. The obtained spatial resolutions in X, Y, and Z coordinates are 1.01~cm, 0.26~cm, and 1.33~cm, respectively (right panel in Fig.~\ref{fig:setups}). These results are very promising. The ability to resolve the annihilation points along the beam ($\sigma$(Y)=0.26) justifies the use of J-PET modules for inertial sensing measurements on 2$^3$S Ps as described in~\cite {MAR20}. Analysis of the complete measured data with both flanges is in progress. A detailed analysis, including the procedure for calibration of the detection modules, description of the analysis algorithm, and final results in terms of achievable spatial resolutions and reconstruction performance of the detectors will be reported in a separate article.
\section{Summary and perspectives}
In this article, we have discussed several research problems that can be studied with the modular detection units of J-PET in the positron beam facility at AML in Trento. The new experimental system at AML can deliver a velocity-moderated, continuous e$^+$ beam. In the next phase, the generation of a monochromatic 2$^3$S Ps beam will be developed. In addition, it is expected that the Ps beam can be produced in a defined quantum state. With the availability of the long-lived 2$^3$S Ps beam, it is planned to use of atomic interferometry to study inertial sensing to measure the gravitational acceleration on Ps. Moreover, the ability to produce Ps in a defined quantum state will enrich studies of entanglement in Ps decays~\cite{BEA19}. These studies will require the registration of multiphotons emitted in Ps decays with good angular and spatial resolution. Modular units based on J-PET technology can be used as potential detectors to perform such studies. A first test with two such modules has already been performed. Preliminary results show that the resolution in spatial coordinates is promising for performing the planned studies. The modular detector units used were developed primarily for tomographic purposes. In the future, a new design with a shorter scintillator length could be considered to meet the specific beam conditions for performing studies with positronium beam at AML. 
\acknowledgments
The authors gratefully acknowledge support from the Foundation for Polish Science through the program TEAM/POIR.04.04.00-00-4204/17; the National Science Centre of Poland through grant no. 2019/35/B/ST2/03562; the Ministry  of Education  and  Science  through  grant  no. SPUB/SP/490528/2021; the SciMat and qLIFE Priority Research Areas budget under the program Excellence Initiative - Research University at the Jagiellonian University, and Jagiellonian University project no. CRP/0641.221.2020. The  authors  also  gratefully  acknowledge  the  support  of  Q@TN,  the  joint laboratory  of  the  University  of  Trento,  FBK-Fondazione  Bruno  Kessler,  INFN-National  Institute  of  Nuclear Physics, and  CNR-National  Research  Council, as  well  as  support  from  the  European  Union's  Horizon  2020 research and innovation programme under the Marie Sklodowska-Curie Grant Agreement No.754496 -FELLINI and Canaletto project for the Executive Programme for Scientific and Technological Cooperation between Italian Republic and the Republic of Poland 2019-2021.

\end{document}